\documentclass[prl,twocolumn]{revtex4}
\usepackage{amssymb}
\usepackage{graphicx}

\begin{document}

\title{Trapped Bose-Einstein condensates with Planck-scale induced
deformation of the energy-momentum dispersion relation }
\author{F. Briscese \thanks{E-mail: fabio.briscese@sbai.uniroma1.it}}

\affiliation{
Istituto Nazionale di Alta Matematica Francesco Severi,\\
Gruppo Nazionale di Fisica Matematica,\\
Citt$\grave{a}$ Universitaria, P.le A. Moro 5, 00185 Rome, Italy.
\\
and\\
DSBAI, Sezione di Matematica, Sapienza Universit$\grave{a}$ di
Roma, Via Antonio Scarpa 16,  00161 Rome,  Italy. }


\begin{abstract}

We show that harmonically trapped Bose-Einstein condensates can be
used to constrain Planck-scale physics. In particular we prove
that a Planck-scale induced deformation of the Minkowski
energy-momentum dispersion relation  $\delta E \simeq \xi_1
mcp/2M_p$ produces a shift in the condensation temperature $T_c$
of about $\Delta T_{c}/T_{c}^{0} \simeq 10^{-6} \xi_1 $ for
typical laboratory conditions. Such a shift allows to bound the
deformation parameter up to $|\xi_1| \lesssim 10^4$. Moreover  we
show that  it is possible to enlarge $\Delta T_{c}/T_{c}^{0}$ and
improve the bound on $\xi_1$ lowering the frequency  of the
harmonic trap. Finally we compare the Planck-scale induced shift
in $T_c$ with similar effects due to interboson interactions and
finite size effects.
\end{abstract}

\maketitle

A general feature of quantum gravitational theories, see
Ref.\cite{smolin} for an introductory reading, is that they imply
a deformation of the standard Minkowski free-particle
energy-momentum dispersion relation at Planck scale, as in the
case of loop quantum gravity \cite{LQG3,LQG4,LQG5,LQG6,LQG7} and
noncommutative geometries \cite{NCG,NCG2,NCG3,NCG4}. Such a
feature opens a possibility to test experimentally  quantum
gravitational  effects \cite{amelinophenomenology}. The free
particle  deformed dispersion relation can be written in a general
form as
\begin{equation}
\begin{array}{ll}
E(p)\equiv E_{0}(p)+\delta E(p,m,M_{P})
\end{array}
\label{deformedenergydensitydefinition0}
\end{equation}
where $E_{0}(p)\equiv \sqrt{p^{2}c^{2}+m^{2}c^{4}}$ is the
Minkowski dispersion relation, $p$ the particle momentum, $m$ its
rest mass, $c$ the speed of light and $M_p$ the Planck mass. The
explicit form of $\delta E(p,m,M_{P})$ depends on the details of
the quantum gravity model used. For simplicity the relation
(\ref{deformedenergydensitydefinition0}) is assumed to be
universal, i.e., is the same for all the elementary particles,
including composite particles such as nucleons or atoms when
internal degrees of freedom are negligible. Since one should
restore the Minkowski dispersion relation well above Planck scale,
one should also require that $\delta E(p,m,M_{P})\rightarrow 0$
when $M_p \rightarrow \infty$. Moreover one would like to preserve
the interpretation of $m$ as the particle rest mass, and therefore
one should impose the additional condition $\delta
E(p=0,m,M_{P})=0$.

In some case the deformation $\delta E$ can be tested in the
ultra-relativistic regime $p \gg m c^2$ by use of astrophysical
data. In the specific case of the deformation $\delta E(p,m,M_{P})
= \eta_1 p^2/2M_p$ such  data could be sensitive to a deformation
parameter $|\eta _{1}| \sim 1$ as discussed  by many authors
\cite{astrophysicalregime,a1,a2,a4,a5}. We mention that a
preliminary analysis of the Fermi Space Telescope data
\cite{fermispacetelescope,f1,f2,f3,f4} is currently  underway.

Very recently \cite{amelino,am1} it was argued the possibility to
constrain the functional form of
(\ref{deformedenergydensitydefinition0}) in the nonrelativistic
regime using ultra-precise cold-atom-recoil-frequency experiments.
Quite generally, in the nonrelativistic regime one can use the
following low momentum ($p \sim 0$) asymptotic expansion of
$\delta E(p,m,M_{P})$ \cite{amelino,am1}
\begin{equation}
\delta E(p,m,M_{P})\simeq \frac{1}{2M_{p}}\left(
\xi_{1}mcp+\xi_{2}p^{2}+\xi_{3}\frac{p^{3}}{mc}\right)
\label{deltaENRparametrization}
\end{equation}
with the real deformation parameter $\xi_{1}$ associated to the
leading term, $\xi_{2}$ to the next leading term, and $\xi_{3}$ to
the next-to-next leading term.

It might be objected that (\ref{deltaENRparametrization}) can be
ruled out for macroscopic objects when $\xi_1 \sim 1$. From
(\ref{deltaENRparametrization}) one has $p^{2}/2m\lesssim \delta
E$ for $ p\lesssim p_{0}\equiv \xi_1 \, c \, m^{2}/M_{P}$ so that
the deformation $\delta E$ dominates over the Minkowski kinetic
term of all momenta up to $p_0$. Standard-model particles with
$m\lesssim 10^{-16}M_{P}$ makes $\delta E$ dominate in the extreme
NR limit $p\lesssim p_{0}\sim \xi_1\,10^{-16}mc$. However, for
macroscopic objects one can easily have $m\sim M_{P}$ and
$p_{0}\sim \xi_1 \, mc$ and therefore the deformation $\delta E$
dominates over Minkowski kinetic term in the entire NR regime. But
this would contradict the familiar dynamics of classical NR bodies
and therefore rule out (\ref{deltaENRparametrization}). This is
what is commonly named ''soccer ball'' problem, see
\cite{soccer,soccer2}. Note, however, that
(\ref{deltaENRparametrization}) is merely the $p \sim 0$
asymptotic expansion of the full deformation $\delta
E(p,m,M_{P})$, and is thus valid for all momenta up to some
$p_{\lambda }$, where $p_{\lambda }$ depends on the explicit
functional form of $\delta E$. For example a deformation

\begin{equation}
\delta E(p,m,M_{P},p_{\lambda })=\xi _{1}\frac{mc\,p}{2M_{P}}\exp
(-p/p_{\lambda })  \label{example}
\end{equation}
behaves as $\delta E=\xi _{1}mc\,p/2M_{P}$ for $p\lesssim
p_{\lambda }$ and $ \delta E\simeq 0$ for $p\gg p_{\lambda }$.
Therefore, to apply (\ref{example}) to macroscopic bodies one
should measure the momenta of extended objects
with $p \lesssim p_{\lambda }$ and this is impossible for sufficiently small $%
p_{\lambda }$ below the lowest measurable momentum for extended
bodies. Since one supposes that this is always the case for
$p_{\lambda }$, the relation (\ref{deltaENRparametrization})
cannot be ruled out for classical macroscopic bodies. We also
emphasize how (\ref{deltaENRparametrization}) is commonly accepted
in the literature \cite{amelino}-\cite{am1}.

In the preceding example $p_\lambda$ acts as a cutoff for the
deformation $\delta E$, above which
Eq.(\ref{deltaENRparametrization}) is no more valid. The value of
the cutoff $p_\lambda$ depends on the specific quantum gravity
model considered, but in principle it can be arbitrarily small.
Therefore, in order to test the validity of
Eq.(\ref{deltaENRparametrization}) for a general class of quantum
gravity models one must capture the effect of the deformation
(\ref{deltaENRparametrization}) at small momentums $p\lesssim
p_\lambda$ for arbitrarily small $p_\lambda$.

For that reason it is meaningful to look at the effect of the
deformation (\ref{deltaENRparametrization}) on the thermodynamic
properties of Bose-Einstein consensates (BECs), which are, in the
case of uniform free condensates, a collection of particles with
$p = 0$ momentum. Moreover, since in the case of harmonically
trapped BECs the single particle ground energy is  $\hbar
\omega/2$ which corresponds to $p = \sqrt{2 \hbar m \omega}$ where
$\omega$ is the frequency of the harmonic trap, one also expects
that Planck-scale effects are strongest for uniform ($p=0$) than
for trapped BECs. As we will show, this expectation is confirmed
by calculations.

In this Letter we calculate the shift in the critical temperature
$T_{c}$ of a harmonically trapped  BEC due to the Planck-scale
induced deformation of the dispersion relation
(\ref{deltaENRparametrization}). The case of non-trapped uniform
BECs in a box has been studied in Ref.\cite{PRDManuel}. Neglecting
interboson interactions and finite size effects in the BEC, it has
been found that that the leading-order deformation in
(\ref{deltaENRparametrization}) produces a shift

\begin{equation}
\Delta T_{c}/T_{c}^{0}\simeq \, 0.1  \, \xi _{1} \, \left(
\frac{m^{2}c}{\hbar M_{P}\,n^{1/3}}\right)   \, \ln[N].
\label{deltaTsuT3}
\end{equation}
where $\Delta T_{c}\equiv T_{c}-T_{c}^{0}$ with $\,T_{c}$ the
critical temperature of the gas with the deformed dispersion
relation, $T_{c}^{0}$ the condensation temperature in the
undeformed Minkowski case, $n$ the boson number density, $m$ the
boson mass, $L^3$ the volume of the box and $N = n L^3$  the total
number of particles. Such a shift  is unexpectedly high if
compared with the strength of the deformation in (\ref
{deltaENRparametrization}) which is of order $\delta E/E \simeq
\xi_1 p/2 c M_{P} \ll \xi_1 m/M_p \sim 10^{-17}$. For a
$_{37}^{87}Rb$ condensate with $N=10^5$ particles, a particle
number density $n\simeq 10^{12}\,cm^{-3}$ and boson mass $m\simeq
150\times 10^{-27}kg$ one finds $\Delta T_{c}/T_{c}^{0}\simeq
5.6\times 10^{-5}\xi _{1}$ \cite{PRDManuel} but, since $\Delta
T_{c}/T_{c}^{0}\propto n^{-1/3}$, such a shift can be even greater
for dilute (smaller $ n$) BECs. In fact, in the case of uniform
BECs, one can reduce $n$ and enlarge the temperature shift without
however making $n$ so small that it invalidates  the thermodynamic
limit. Alternatively one can increase the number of particles but,
since the dependence on $N$ is logarithmic, one should consider
huge $N$ to considerably rise the temperature shift. Moreover, the
effect of the next-to-leading-term deformation in
(\ref{deltaENRparametrization}) was found to be $\Delta
T_{c}/T_{c}^{0}=\xi _{2}m/M_{P}$ which   is extremely small.

The relevant result (\ref{deltaTsuT3}) encourages a further
investigation of the problem. In facts BECs are produced in the
laboratory in laser-cooled, magnetically-trapped ultra-cold
bosonic clouds \cite{dellano2}, e.g. $_{37}^{87}Rb$
\cite{Ander},$\ _{3}^{7}Li$ \cite{Bradley}, $ _{11}^{23}Na$
\cite{Davis}, $_{1}^{1}H$ \cite{Fried}, $_{37}^{87}Rb$ \cite
{Cornish}, $_{2}^{4}He$ \cite{Pereira}, $_{19}^{41}K$
\cite{Mondugno}, $ _{55}^{133}Cs$ \cite{Grimm}, $_{70}^{174}{Yb}$
\cite{Takasu03} and $ _{24}^{52}Cr$ \cite{Griesmaier}, $^{84}$Sr
\cite{sr}, $^{168}$Er \cite{er}. Therefore, to deal with
laboratory measurements of $\Delta T_{c}/T_{c}^{0}$ one should
generalize (\ref{deltaTsuT3}) to the case of  trapped BECs.

In the following  we will show that in the case of harmonically
trapped BECs one has a Planck-induced temperature shift $\Delta
T_{c}/T_{c}^{0} \sim 10^{-6} \xi_1$ for the leading order
deformation in (\ref{deltaENRparametrization}). This allows to
bound the deformation parameter up to $|\xi_1| \lesssim 10^4$.
Such a bound is four orders of magnitude above the best bound
$|\xi_1| \lesssim 1$ obtained with ultra-precise
cold-atom-recoil-frequency experiments \cite{amelino,am1}. However
it is notable the possibility to test Planck-scale effects with
BECs. We will also discuss the possibility to improve such a bound
considering  BECs with low values of the harmonic trap frequency
$\omega$. Moreover we will compare Planck-scale effects with the
shift in the condensation temperature due to finite size effects
and to interboson interactions. Finally we will show that the
effect of the next to leading order term in
Eq.(\ref{deltaENRparametrization}) gives an extremely small
temperature shift $\Delta T_{c}/T_{c}^{0} = \xi_2 m/M_p$ and
therefore the deformation parameter $\xi_2$ cannot be bounded
significantly.

Let us consider a system composed of  N bosons trapped in an
external spherically symmetric harmonic potential $V_{ext} = m
\omega^2 r^2/2$.  We  define $\varphi_0(r) \equiv
(m\omega/\pi\hbar)^{3/4} \exp[-V_{ext}(r)/\omega\hbar]$ that is
the wave function of the ground state of the single particle
quantum harmonic oscillator. The  condensate distribution at zero
temperature  is that of N identical bosons in the ground state of
a harmonic oscillator $n_c(r) = N |\varphi_0(r)|^2 = N
\left(\sqrt{\pi} a_0 \right)^{-3} \exp[-m \omega r^2/\hbar]$
\cite{dalfovo,parkinswalls,burnett}. The typical size of the
support of $\varphi_0$  is $a_0= \sqrt{\hbar/m\omega}$ and
represents the size of the condensate at $T=0$ temperature. The
typical size of the thermal cloud at finite temperatures $k_B T
\gg \hbar \omega$ is $R_T = a_0 \sqrt{2 \pi k_B T/\hbar\omega}\gg
a_0$ \cite{dalfovo,parkinswalls,burnett}. We also remember that in
the case of a harmonically trapped BEC the thermodynamic limit is
given by $N \rightarrow \infty$ and $\omega \rightarrow 0$ with $N
\omega^3$  finite. Finally we remember that the condensation
temperature of the ideal harmonically trapped BEC with the usual
Minkowski dispersion relation ($\xi_1=\xi_2=\xi_3=0$) in the
semiclssical limit $k_B T \gg \hbar \omega$ is given by $k_B T^0_c
= \hbar \omega \left(N/\zeta(3)\right)^{1/3}\gg \hbar \omega$.

The semiclassical energy in phase space of nonrelativistic
particles in presence of an external potential is $E(p,r) = m c^2
+ p^2/2m + \delta E(p) + V_{ext}(r)$, see
\cite{dalfovo,parkinswalls,burnett}. By means of this expression
the number of bosons in thermal equilibrium $N_{th}$ is given by
\cite{dalfovo,parkinswalls,burnett,cong}

\begin{equation}\label{semiclassicalNth2}
N_{th}= \int \frac{d^3x d^3p}{(2\pi \hbar)^3}
\frac{1}{\exp[\beta(E(p,r)-\mu)]-1}.
\end{equation}

The expression (\ref{semiclassicalNth2}) allows to calculate the
shift in the condensation temperature due to the deformation
(\ref{deltaENRparametrization}) as follows. We parameterize the
energy deformation (\ref{deltaENRparametrization}) in the general
form $\delta E(p) = \alpha f(p)$ where $\alpha \ll 1$ is a
dimensionless deformation parameter. With such a definition the
condensation temperature $T_{c}(\alpha)$ is a function of
$\alpha$. It is obtained by extracting $T_{c}(\alpha)$ from
Eq.(\ref{semiclassicalNth2}) with the substitutions $N_{th} =N$
and $\mu = m c^2$, that is

\begin{equation} \label{numericaltc}
\frac{\pi \hbar^3 N}{2} = \int  \frac{dr dp \, r^2
\,p^2}{\exp[\frac{p^2/2m + \alpha f(p)+ V_{ext}(r)}{k_B
T_c(\alpha)}]-1}.
\end{equation}
Since the lhs of (\ref{numericaltc}) is independent of $\alpha$
one has $\partial _{\alpha}N=0$ and after some algebra one obtains
\[
\frac{\partial _{\alpha}T_{c}(\alpha)}{T_{c}(\alpha)}= \int dr dp
\, r^2 p^2 f(p) g(p,r,\alpha)p^{2}/\int dr dp \, r^2 p^2 \times
\]
\begin{equation}
\left[ p^2/2m + \alpha \, f(p)+ V_{ext}(r) \right] g(p,r,\alpha)
\label{deltaTsuT1}
\end{equation}
where
\begin{equation}
g(p,r,\alpha)\equiv  \frac{\exp \left[
\frac{p^2/2m+\alpha\,f(p)+V_{ext}(r)}{k_{B}T_{c}(\alpha)}\right]}{\left[
\exp \left[
\frac{p^2/2m+\alpha\,f(p)+V_{ext}(r)}{k_{B}T_{c}(\alpha)} \right]
-1\right]^{2}}.
\end{equation}
We can use this expression  to calculate the shift in $T_{c}$.
Since $\alpha \ll 1$, one has
\begin{equation}
\frac{\Delta
T_{c}}{T_{c}^{0}}=\frac{T_{c}(\alpha)-T_{c}(0)}{T_{c}(0)}\simeq
\alpha \left( \frac{\partial
_{\alpha}T_{c}(\alpha)}{T_{c}(\alpha)}\right) _{|_{\alpha=0}}
\label{temperatureshift}
\end{equation}
and the last term is evaluated by use of Eq.(\ref{deltaTsuT1})
remembering that $T_c(0) = T_c^0$.

We are first interested in the leading term of the deformation
(\ref{deltaENRparametrization}), that is $\delta
E=\xi_{1}mcp/2M_{P}$ which corresponds to $\alpha=\xi _{1}m/2M_{P}
\ll 1$ and $f(p)=cp$. In this case the resulting temperature shift
is
\begin{equation}
\Delta T_{c}/T_{c}^{0}\simeq 0.3 \, \xi _{1} \left(m/M_p \right)
\sqrt{ m c^{2}/\hbar \omega N^{1/3}}. \label{HarmonicdeltaTsuT}
\end{equation}
and is finite in the thermodynamic limit, since it depends on
$\omega$ and $N$ through $N \omega^3$. For  a $_{37}^{87}Rb$
condensate  in an harmonic spherical potential trap with typical
values of  $N \sim 10^{5}$ and frequency $\omega \sim 10 \, Hz$
\cite{dalfovo,parkinswalls,burnett}, Eq.(\ref{HarmonicdeltaTsuT})
gives $\Delta T_{c}/T_{c}^{0} \simeq 10^{-6} \xi_1$. This is about
two order of magnitude smaller than the non-trapped case $\sim
10^{-4} \xi_1$ \cite{PRDManuel}, but still extremely large if
compared with the strength of deformation in
Eq.(\ref{deltaENRparametrization}) $\delta E/E \ll 10^{-17}$. This
estimation of the Planck-induced $\Delta T_{c}/T_{c}^{0}$ should
be compared with current experimental precision in $T_{c}$
measurements. In high-precision measurements of $\Delta
T_{c}/T_{c}^{0}$\ in $_{19}^{39}K$ \cite{condensatePRL} due to
interboson interactions, the order of magnitude is $\Delta
T_{c}/T_{c}^{0}\simeq 5\times 10^{-2}$ with a $\sigma_{\Delta
T_{c}/T_{c}^{0}} \lesssim 10^{-2}$. This allows to bound the
deformation parameter up to $|\xi_1|$ $\lesssim 10^{4}$, which is
about four orders of magnitude above the best estimation $|\xi_1|$
$\lesssim 1$ obtained in cold-recoil-frequency experiments
\cite{amelino,am1}, but is nevertheless notable.

One can however look for physical situations in which  the
temperature shift is enlarged and the bound on $\xi_1$ improved.
In Ref.\cite{PRDManuel} it was argued that in the case of uniform
BECs the temperature shift (\ref{deltaTsuT3}) is enlarged in
dilute condensates since $\Delta T_{c}/T_{c}^{0} \propto
n^{-1/3}$. Let us analyze the case of harmonically trapped
condensates. The central density of an ideal trapped BEC is
$n_c(0) = N/(\sqrt{\pi} a_0)^3 \propto N \omega^{3/2}$ and it goes
to infinitum in the thermodynamic limit. Moreover $n_c(0)$ is
strongly suppressed in non-ideal BECs due to interactions (quantum
depletion), so it is not  useful to parameterize  the temperature
shift. We can define the following number density  $n_T \equiv
N/R^3_T = (N \omega^3/c^3) (mc^2/2\pi k_B T)^{3/2}$ which is the
ratio between the number of bosons and the volume occupied by the
thermal cloud and is well defined at finite temperatures in the
thermodynamic limit. With such a definition one has $\Delta
T_{c}/T_{c}^{0} \propto n_T^{-1/6}$, therefore in the case of
trapped condensates one can enlarge $\Delta T_{c}/T_{c}^{0}$ by
reducing $n_T$ which in turns implies to reduce $N \omega^3$.
Because of the $-1/6$ exponent the dependence of the temperature
shift on $n_T$ is very smooth and it seems that one cannot
significantly enlarge $\Delta T_{c}/T_{c}^{0}$ reducing $n_T$.
However one has $\Delta T_{c}/T_{c}^{0} \propto \omega^{-1/2}
N^{-1/6}$ from which one see that one can enlarge $\Delta
T_{c}/T_{c}^{0}$ lowering $\omega$. Also note that the dependence
of the temperature shift with respect to N is very smooth and one
cannot lower $N$ without loosing the validity of the thermodynamic
limit, therefore one cannot rise $\Delta T_{c}/T_{c}^{0}$ by
lowering $N$. In conclusion in the case of harmonically trapped
condensates one can rise Planck-scale effects on $T_c$ and
therefore improve the bounds on $\xi_1$ considering BECs with low
$n_T$, which corresponds to low frequencies $\omega$ and low
condensation temperatures $T_c$.

We  stress that an improvement of the precision in $T_c$
measurement would also allow to improve the bounds on $\xi_1$.
However, even in the ideal situation of an extremely precise
measurement of $T_c$, in order to constrain Planck-scale effects
with condensation temperature measurements one has to deal with
the effect of interboson interactions and with finite size
effects, which both affect $T_c$.

The shift in $T_c$ due to interboson interactions is $\Delta
T_{c}/T_{c}^{0} \simeq - 3.426 \, a/\lambda_T$ where $\lambda_T =
\hbar/\sqrt{2 \pi m k_B T^0_c}$ is the thermal wavelength at
temperature $T_c^0$ and $a$ is the s-wave scattering length
\cite{dalfovo,parkinswalls,burnett}. Such a shift can be as large
as $ 4 \times 10^{-2}$ for $a \sim 10^2 \, a_0$ and $\lambda_T
\sim 10^4 \, \, a_0$ where $a_0$ is the Bohr radius, and it has
been measured in $^{39}_{19}K$ \cite{condensatePRL} where the data
are well fitted by a second order polynomial $\Delta
T_{c}/T_{c}^{0} \simeq b_1 (a/\lambda_T) +  b_2 (a/\lambda_T)^2$
with $b_1 = -3.5 \pm 0.3$ and $b_2 = 46 \pm 5$, the second term
being due to beyond-MF effects. Therefore in such a case the shift
in $T_c$ due to interboson interaction is about four order of
magnitude greater than the Planck-scale induced shift for $\xi_1
\sim 1$. This problem may be solved in principle, just tuning the
interaction coupling by Feshbach resonances to very small values
of $a$, therefore reducing the contribution of interboson
interactions on $\Delta T_{c}/T_{c}^{0}$ below the Planck-scale
induced shift $\sim 10^{-6} \xi_1$. However this may cause some
practical difficulty. In fact, interboson interactions drive
bosons to thermodynamic equilibrium and any experiment is limited
by the fact that for extremely small interactions equilibrium is
never reached.

Finite size effects are due to the finiteness of $N$ in comparison
with the thermodynamic limit $N \rightarrow \infty$. The shift in
the condensation temperature due to finite size effects is
estimated as $\Delta T_{c}/T_{c}^{0} \simeq -0.73 \, N^{-1/3}$
\cite{grossman,ketterle,kirsten} and typically it is of about
$10^{-2}$. Therefore, finite size effects are about four order of
magnitude larger than the expected Planck-scale induced shift
$\sim 10^{-6}$ for $\xi_1 \sim 1$ and typical $N \sim 10^5 -
10^{6}$. We stress that finite size effects are technologically
impossible to be tuned below Planck-scale induced effects at least
for existing condensates, since this would require huge $N \gtrsim
10^{30}$ corresponding to extremely small shifts $\Delta
T_{c}/T_{c}^{0}\lesssim 10^{-10}$. Thus,  in order to measure a
deformation parameter $|\xi_1| $ one should predict the finite
size contributions to $\Delta T_{c}/T_{c}^{0}$ for typical
laboratory conditions $N \sim 10^5 - 10^{6}$ with an accuracy up
to $\sim 10^{-6} |\xi_1|$. However, experimental strategies allows
to eliminate finite size effects from  $\Delta T_{c}/T_{c}^{0}$
measures as in \cite{condensatePRL}. In fact for each measurement
series at a given $a$ and $\lambda_T$, a reference measurement is
taken with a small $a/\lambda_T \sim 0.005$, same $\omega$ and
very similar $N$, hence very similar $\lambda_T$. Thus one can
eliminate all $a$-independent systematic errors that usually
affect absolute measurements of $T_c$ including uncertainties in
the absolute calibration of $N$ and $\omega$ as well as the  shift
due to finite size effects.

We remark that, even if we have traced a possible approach to the
problem, dealing with interboson interaction and finite size
effects in real experiments is a strong matter which deserves
further investigation that will be presented elsewhere
\cite{briscese2}.

Finally we consider  the next to leading order deformation in
Eq.(\ref{deltaENRparametrization}) which is $\delta E = \xi_2
p^2/2M_p$ and corresponds to $\alpha = \xi_2 m/M_p$ and $f(p) =
p^2/2m$. With such a deformation Eq.(\ref{temperatureshift}) gives
$\Delta T_{c}/T_{c}^{0} \simeq \xi_2 m/M_p$ which  is extremely
small of order $\sim 10^{-17}$ and does not allow to constrain
significantly the next to leading order parameter $\xi_2$.

In conclusion, the main goal of this Letter has been to show that
trapped BECs can be used  to constrain Planck-scale physics. In
particular we have shown that the leading order dispersion
relation deformation defined in Eq.(\ref{deltaENRparametrization})
produces a shift in the condensation temperature of about $\Delta
T_{c}/T_{c}^{0} \simeq 10^{-6} \xi_1 $ for typical laboratory
conditions and such a shift allows to bound the deformation
parameter up to $|\xi_1| \lesssim 10^4$. Moreover  we have
discussed how it is possible to enlarge such a shift and improve
the bound on $\xi_1$ lowering the frequency $\omega$ of the
harmonic trap. Finally we have compared the Planck-scale induced
shift with similar effects due to finite size and interboson
interactions.

During the edition of this Letter  a similar analysis of the
Planck-scale induced temperature shift in trapped BECS has been
realized \cite{castellanos1,castellanos2}. However, in addition to
such analysis, here we give a stronger focus to the the
possibility of planning specific experiments  that might provide
phenomenological constraints on Planck-scale physics.

\textbf{Acknowledgements}: I would like to thank M. de Llano, M.
Grether  and G. Amelino-Camelia for useful discussions during the
edition of this Letter. F. Briscese is a Marie Curie fellow of the
Istituto Nazionale di Alta Matematica Francesco Severi.

\end{document}